\documentclass[journal]{IEEEtran}%

\usepackage[printonlyused]{acronym} 
\acrodef{ADC}{analog-to-digital converter}
\acrodef{ADM}{add-drop multiplexer}
\acrodef{AOM}{acusto-optic modulator}
\acrodef{ASE}{amplified spontaneous emission}
\acrodef{AIR}{achievable information rate}
\acrodef{AKNS}{Ablowitz-Kaup-Newell-Segur}
\acrodef{AWG}{arbitrary waveform generator}
\acrodef{AWGN}{additive white Gaussian noise}
\acrodef{B2B}{back-to-back}
\acrodef{BER}{bit error rate}
\acrodef{BPD}{balanced photodetector}
\acrodef{BPS}{blind phase search}
\acrodef{CFO}{carrier frequency offset}
\acrodef{DAC}{digital-to-analog converter}
\acrodef{DC}{direct current}
\acrodef{DCF}{dispersion-compensating fiber}
\acrodef{DBP}{digital back-propagation}
\acrodef{DFT}{discrete Fourier transform}
\acrodef{DP-NFDM}{dual-polarization nonlinear frequency division multiplexing}
\acrodef{DSO}{digital storage oscilloscope}
\acrodef{DSP}{digital signal processing}
\acrodef{DT}{Darboux transform}
\acrodef{EDC}{electrical dispersion compensation}
\acrodef{EDFA}{erbium-doped fiber amplifier}
\acrodef{ENOB}{effective number of bits}
\acrodef{EVM}{error vector magnitude}
\acrodef{FEC}{forward error correction}
\acrodef{FFT}{fast Fourier transform}
\acrodef{FL}{fiber laser}
\acrodef{FWHM}{full-width half-maximum}
\acrodef{FWM}{four-wave mixing}
\acrodef{GLM}{Gelfand-Levitan-Marchenko}
\acrodef{GLME}{Gelfand-Levitan-Marchenko equation}
\acrodef{GVD}{group velocity dispersion}
\acrodef{HD-FEC}{hard-decision forward error correction}
\acrodef{I}{in-phase}
\acrodef{IFFT}{inverse fast Fourier transform}
\acrodef{INFT}{inverse nonlinear Fourier transform}
\acrodef{ISI}{inter-symbol interference}
\acrodef{ISO}{isolator}
\acrodef{IST}{inverse scattering transform}
\acrodef{IVP}{initial value problem}
\acrodef{LO}{local oscillator}
\acrodef{LPA}{lossless path-averaged}
\acrodef{ME}{Manakov equation}
\acrodef{MS}{Manakov system}
\acrodef{MAP}{maximum a posteriori}
\acrodef{ML}{maximum likelihood}
\acrodef{MZM}{Mach-Zehnder modulator}
\acrodef{MZSP}{Manakov-Zakharov-Shabat spectral problem}
\acrodef{NFDM}{nonlinear frequency division multiplexing}
\acrodef{NFT}{nonlinear Fourier transform}
\acrodef{NCG}{Nystr{\"o}m conjugate gradient}
\acrodef{NIS}{nonlinear inverse synthesis}
\acrodef{NLSE}{nonlinear Schr\"odinger equation}
\acrodef{NMSE}{normalized mean squared error}
\acrodef{OBPF}{optical band pass filter}
\acrodef{ODE}{ordinary differential equation}
\acrodef{OFDM}{orthogonal frequency-division multiplexing}
\acrodef{OPC}{optical phase conjugation}
\acrodef{OSNR}{optical signal-to-noise ratio}
\acrodef{PAPR}{peak-to-average power ratio}
\acrodef{PC}{polarization controller}
\acrodef{PDE}{partial differential equation}
\acrodef{PMD}{polarization mode dispersion}
\acrodef{ppm}{parts-per-million}
\acrodef{PRBS}{pseudo-random bit sequence}
\acrodef{PSK}{phase-shift keying}
\acrodef{Q}{quadrature}
\acrodef{QAM}{quadrature amplitude modulation}
\acrodef{QPSK}{quadrature phase-shift keying}
\acrodef{SDM}{space-division multiplexing}
\acrodef{SMF}{single-mode fiber}
\acrodef{SNR}{signal-to-noise ratio}
\acrodef{S/P}{serial-to-parallel}
\acrodef{SPM}{self-phase modulation}
\acrodef{SSFM}{split-step Fourier method}
\acrodef{WDM}{wavelength-division multiplexing}
\acrodef{XPM}{cross-phase modulation}
\acrodef{ZSP}{Zakharov-Shabat spectral problem}
\acrodef{Z-S}{Zakharov-Shabat}

\usepackage{xcolor}

\usepackage{amsthm}
\theoremstyle{plain}

  \theoremstyle{plain}
  
\providecommand{\corollaryname}{Corollary}
\providecommand{\theoremname}{Theorem}

\usepackage{cite}

\ifCLASSINFOpdf
  \usepackage[pdftex]{graphicx}
\else
   \usepackage[dvips]{graphicx}
   \DeclareGraphicsExtensions{.eps}
\fi
\usepackage[cmex10]{amsmath}
\usepackage{url}

\usepackage{hyperref}
\usepackage[OT4]{fontenc}
\usepackage{soul}

\input{usrpackages}
\usepackage{xargs}        
\usepackage{etoolbox}     
\usepackage{physics}      
\usepackage{amsfonts}     
                          
\newcommand{\RR}{\mathbb{R}}                
\newcommand{\CC}{\mathbb{C}}                
\newcommand{\iunit}{j}                      
\newcommand{\vecbold}[1]{                   
  \boldsymbol{\mathbf{#1}}}       

\newcommand{\kk}{j}                         
\newcommand{\ii}{i}                         
\newcommand{\nEig}{N_{DS}}                       

\newcommand{\dispersion}{\beta_2}           
\newcommand{\nonlinfact}{\gamma}            

\newcommand{\ttm}{\tau}                        
\newcommand{\ssp}{\ell}                        
\newcommandx{\flds}[2][1, 2]{                  
    \des[#2]{E}{#1}}
\newcommandx{\fldsrc}[3][1 ,2, 3]{             
  \flds[#1][#3](\ttm_{#2}}                     
\newcommandx{\fld}[3][1, 2, 3]{                
\fldsrc[#1][#2][#3] )}                         
\newcommandx{\fldl}[3][1, 2, 3]{               
\fldsrc[#1][#2][#3], \ssp) }                   

\newcommand{\nttm}{t}                       
\newcommand{\nssp}{z}                       
\newcommand{\To}{T_{0}}                     
\newcommand{\nftL}{\mathcal{L}}             

\newcommand{\nfldletter}{q}
\newcommandx{\nflds}[2][1, 2]{              
    \des[#2]{\nfldletter}{#1}}
\newcommandx{\nfldsrc}[3][1 ,2, 3]{         
  \nflds[#1][#3](\nttm_{#2}}                
\newcommandx{\nfld}[3][1, 2, 3]{            
  \nfldsrc[#1][#2][#3] )}                   
\newcommandx{\nfldl}[3][1, 2, 3]{           
  \nfldsrc[#1][#2][#3], \nssp )}            

\newcommandx{\nftquantity}[5][1, 2, 3, 4, 5]{       
  \des[#4]{#1}{#3}(\eig[#2][#5]}                    
                                                    
\newcommandx{\eig}[2][1, 2]{#2{\lambda}_{#1}}  

\newcommand{\nftasrc}{a}                       
\newcommandx{\nfta}[2][1, 2]{#2{\nftasrc}(\eig[#1])}         
\newcommandx{\nftaderiv}[2][1, 2]{#2{\nftasrc}'(\eig[#1])}   

\newcommandx{\nftbsrc}[3][1= ,2=, 3=]{       
  \nftquantity[b][#1][#2][#3]}                                              
\newcommandx{\nftb}[3][1= ,2=, 3=]{          
  \nftbsrc[#1][#2][#3] ) }     
\newcommandx{\nftbl}[3][1= ,2=, 3=]{         
  \nftbsrc[#1][#2][#3], \nssp) }     
  
\newcommandx{\cnftsrc}[2][1= ,2=]{          
  \nftquantity[Q_c][][#1][#2] }
\newcommandx{\cnft}[2][1= ,2=]{             
  \cnftsrc[#1][#2] ) }
\newcommandx{\cnftl}[2][1= ,2=]{            
  \cnftsrc[#1][#2], \nssp) }             
  
\newcommandx{\dnftsrc}[3][1=, 2=, 3=]{      
  \nftquantity[Q_d][#1][#2][#3] }  
\newcommandx{\dnft}[3][1=, 2=, 3=]{         
  \dnftsrc[#1][#2][#3] ) }
\newcommandx{\dnftl}[3][1=, 2=, 3=]{        
  \dnftsrc[#1][#2][#3], \nssp) }
  
\newcommandx{\nftInvH}[2][1=, 2=]{e^{4\iunit\eig[#1]^2#2}}   
\newcommandx{\nftInvHtext}[2][1=, 2=]{\exp(4\iunit\eig[#1]^2#2)}  
    
\newcommandx{\nftbconj}[2][1=, 2=]{        
\nftb[#1][#2][\bar]}

\newcommand{\estoperator}{\widehat}        

\newcommandx{\nftbest}[2][1=, 2=]{         
    \nftbsrc[#1][#2]
    [\estoperator][\estoperator] ) }
\newcommandx{\cnftest}[1][1=]{             
    \cnftsrc[#1]
    [\estoperator][\estoperator] ) }
\newcommandx{\dnftest}[2][1=, 2=]{         
    \dnftsrc[#1][#2]
    [\estoperator][\estoperator] ) }
    
\newcommand{\matL}{\boldsymbol{L}}

\newcommandx{\eigf}[2][1, 2]{\des[#2]{v}{#1}} 

\newcommandx{\jostquantity}[3][1, 2, 3]{    
  \des[#3]{#1}{#2}}                         
\newcommand{\jostconjoperator}{\bar}            

\newcommandx{\jostps}[2][1 ,2]{             
  \jostquantity[\psi][#1][#2]}              
\newcommandx{\jostpsrc}[2][1 ,2]{           
  \jostps[#1][#2](\nttm}              
\newcommandx{\jostp}[2][1 ,2]{              
  \jostpsrc[#1][#2] ) }  
\newcommandx{\jostpl}[3][1 ,2, 3]{          
  \jostpsrc[#1][#3], \eig[#2] ) }
\newcommandx{\jostpll}[3][1 ,2, 3]{         
  \jostpsrc[#1][#3], \eig[#2]; \nflds ) }

\newcommandx{\jostpconjs}[2][1, 2]{         
\jostps[#1][\jostconjoperator]}
\newcommandx{\jostpconj}[2][1, 2]{          
\jostp[#1][\jostconjoperator]}
\newcommandx{\jostpconjl}[2][1, 2]{         
\jostpl[#1][#2][\jostconjoperator]}

\newcommandx{\jostns}[2][1 ,2]{             
  \jostquantity[\phi][#1][#2]}              
\newcommandx{\jostnsrc}[2][1 ,2]{           
  \jostns[#1][#2](\nttm}              
\newcommandx{\jostn}[2][1 ,2]{              
  \jostnsrc[#1][#2] ) }  
\newcommandx{\jostnl}[3][1 ,2, 3]{         
  \jostnsrc[#1][#3], \eig[#2] ) }
\newcommandx{\jostnll}[3][1 ,2, 3]{        
  \jostnsrc[#1][#3], \eig[#2]; \nflds ) }

\newcommandx{\jostnconjs}[2][1, 2]{          
\jostns[#1][\jostconjoperator]}
\newcommandx{\jostnconj}[2][1, 2]{          
\jostn[#1][\jostconjoperator]}
\newcommandx{\jostnconjl}[2][1, 2]{         
\jostnl[#1][#2][\jostconjoperator]}

\newcommand{\darboperator}{\widehat}    
\newcommandx{\darbauxsols}[2][1, 2]{  
  \des[#2]{\nu}{#1}}               
\newcommandx{\darbauxsol}[3][1, 2, 3]{  
  \darbauxsols[#1][#3](                 
    \ifstrequal{#2}{}{\nttm}{#2}
  )}
\newcommandx{\darbauxsoll}[4][1, 2, 3, 4]{  
  \darbauxsols[#1][#4](                     
    \ifstrequal{#3}{}{\nttm}{#3}, 
  \eig[#2])}

\newcommandx{\nflddarbs}[1][1]{       
  \nflds[#1][\darboperator] }
\newcommandx{\nflddarb}[2][1, 2]{       
  \nfld[#1][#2][\darboperator] }
\newcommandx{\eigfdarb}[1][1]{          
\eigf[#1][\darboperator] }

\newcommand{\predarboperator}{\widetilde}

\newcommandx{\cnftpredarb}[1][1=]{      
  \cnft[#1][\predarboperator] }
\newcommandx{\nfldpredarbs}[2][1, 2]{   
\nflds[#1][\predarboperator]}      
\newcommandx{\nfldpredarb}[2][1, 2]{    
\nfld[#1][#2][\predarboperator]}       
\newcommandx{\nfldpredarbl}[2][1, 2]{   
\nfldl[#1][#2][\predarboperator]}      
\newcommandx{\jostnpredarbll}[3][1 ,2, 3]{        
  \jostnsrc[#1][#3], \eig[#2]; \nflds[][\predarboperator] ) }
\newcommandx{\jostppredarbll}[3][1 ,2, 3]{        
  \jostpsrc[#1][#3], \eig[#2]; \nflds[][\predarboperator] ) }
  
\newcommand\des[3][]{\desaux{#2}{#3}#2_\relax{#1}}
\def\desaux#1#2#3_#4\relax#5{%
  \ifx\relax#4\relax
    \ifx\relax#2\relax
      \vecbold{#5#1}
    \else
      #5#1_{#2}
    \fi
  \else
    \ifx\relax#2\relax
      \vecbold{#5#3_{\stripus#4}}
    \else
      #5#3_{\stripus#4,#2}
    \fi
  \fi
}
\def\stripus#1_{#1}

\renewcommand{\ul}{}

\begin{document}
%
\title{Dual-polarization NFDM transmission with continuous and discrete spectral modulation}
%
%
%


\author{F. Da Ros, S. Civelli, S. Gaiarin, E.P. da Silva, N. De Renzis, M. Secondini, and D. Zibar	
\thanks{Manuscript received October 18th, 2018; }%
\thanks{F. Da Ros, S. Gaiarin, N. De Renzis, and D. Zibar  are with the Department of Photonics Engineering, Technical University of Denmark, Kongens Lyngby, 2800 Denmark, e-mail: \{fdro,simga,nidre,dazi\}@fotonik.dtu.dk}
\thanks{S. Civelli and M. Secondini are with  the TeCIP Institute, Scuola Superiore Sant'Anna, Pisa, Italy, email: \{stella.civelli,marco.secondini\}@santannapisa.it}%
\thanks{E.P. da Silva was with the Department of Photonics Engineering, Technical University of Denmark, Kongens Lyngby, 2800 Denmark. He is now with the Department of Electrical Engineering of the Federal University of Campina Grande (UFCG), Paraíba, Brazil, email:edson.silva@dee.ufcg.edu.br.}%
\thanks{F. Da Ros, S. Civelli and S. Gaiarin equally contributed to this work.}%
}
%
%

\markboth{Submitted to IEEE Journal of Lightwave Technology}%
{}
\markboth{Submitted to IEEE Journal of Lightwave Technology}{}%
%



\maketitle

\begin{abstract}
Nonlinear distortion experienced by signals during their propagation through optical fibers strongly limits the throughput of optical communication systems. Recently, a strong research focus has been dedicated to nonlinearity mitigation and compensation techniques. At the same time, a more disruptive approach, the \ac{NFT}, aims at designing signaling schemes more suited to the nonlinear fiber channel. In a short period, impressive results have been reported by modulating either the continuous spectrum or the discrete spectrum. Additionally, very recent works further introduced the opportunity to modulate both spectra for single polarization transmission. Here, we extend the joint modulation scheme to dual-polarization transmission by introducing the  framework to construct a dual-polarization optical signal with the desired continuous and discrete spectra. After a brief analysis of the numerical algorithms used to implement the proposed scheme, the first experimental demonstration of dual-polarization joint \ac{NFDM} modulation is reported for up to 3200~km of low-loss transmission fiber. The proposed dual-polarization joint modulation schemes enables to exploit all the degrees of freedom for modulation (both polarizations and both spectra) provided by a \ac{SMF}.
\end{abstract}


%

\section{Introduction}
\label{sec:Intro}
\IEEEPARstart{O}{ptical} communication systems have experienced an impressive growth over the past few decades with ever increasing transmission rates. Such a growth has been the results of key enabling technologies that  have allowed to counteract the several physical effects hindering the transmission. As linear effects such as loss and dispersion can be dealt with in the telecommunication C-band, one of the current limiting factor preventing to further enhance the throughput of \acp{SMF} is the impact of Kerr nonlinearity. 
Whereas a significant research effort has been devoted to counteract nonlinear distortion experienced by the signals during propagation, with solutions presented both in the optical and digital domain~\cite{EllisAOP17}, no clear practical solution has yet been devised.
Over the past few years, a theoretical approach, which had led to soliton-based communication, has been rediscovered~\cite{YousefiTransInfTh14,TuritsynOptica17}. Soliton-based communication was developed in the 1980's~\cite{HasegawaAPL73,MollenauerOL88,NakazawaEL91},\ul{ and, in particular, eigenvalue based communication was first attempted in~}\cite{Hasegawa93},\ul{ before the advent of }\ac{WDM},\ul{ and it was quickly abandoned due to its low spectral efficiency, challenges with fiber loss, noise and soliton-soliton interaction and the lack of coherent transceivers enabling access to the full electrical field}. Recently, however, a generalization of the mathematical theory behind soliton communication, i.e., the \ac{IST}, has been exploited to devise a new approach to transmit over \acfp{SMF}. The \ac{IST}, re-branded as \acf{NFT} within the optical communication community, provides a transformation that enables to effectively linearize the \ac{NLSE} describing the optical wave propagation through a \ac{SMF}. \textcolor{black}{By using such a transformation, the impact of group velocity dispersion and Kerr nonlinearity can be constructively taken into account to design a novel signaling system that aims at not being limited by signal-signal nonlinear interaction.} 
Furthermore, this transformation \textcolor{black}{may provide more flexible modulation techniques} compared to standard coherent approaches, as it associates two spectral quantities to one time-domain waveform: a continuous spectrum corresponding to dispersive waves and a discrete spectrum representing the solitonic solutions~\cite{YousefiTransInfTh14}. Since the \ac{NFT} theory strictly requires loss-less, noise-less, and dispersion-slope free transmission, a significant research effort has been devoted into addressing these requirements both numerically, e.g., by using the \ac{LPA} approximation~\cite{TuritsynPR12,LeJLT15,LeOE15}, and experimentally, e.g., using distributed Raman amplification~\cite{GaiarinPTL18,GuiJLT17,GejslerITG18}.
This has yield several impressive demonstrations by encoding information in either the continuous~\cite{LeJLT16} \ul{(up to 3.3~Gb/s over 7344~km)}, or the discrete spectrum~\cite{ArefECOC15} \ul{(4~Gb/s over 640~km)} \cite{BuelowOFC16} \ul{(4~Gb/s over 1600~km)} \cite{GejslerITG18} \ul{(4~Gb/s over 4900~km)} and \cite{DongPTL15} \ul{(1.5~Gb/s over 1800~km)}, finally leading to a recent report of a joint continuous and discrete spectral modulation~\cite{ArefJLT18,LeNatPhot17} \ul{(up to 64~Gb/s over 976~km)}.
In parallel, the use of \acp{SMF} for transmission provides an additional degree of freedom to increase the transmission throughput by exploiting two orthogonal field polarizations. The \ac{NFT} for the \ac{NLSE} has therefore been extended to the \ac{MS}~\cite{ManakovJETP74}, where two signal polarizations are transmitted under the assumption that the state-of-polarization varies fast enough to be averaged over the nonlinear and dispersion lengths, and polarization mode dispersion can be neglected. The theory has been applied to numerical and experimental demonstrations of dual-polarization \ac{NFT}-based transmission using either the continuous~\cite{GoossensOE17,CivelliOE18pol,GuiCLEO18} or the discrete spectrum~\cite{GaiarinOptica17,GaiarinPTL18}. A dual-polarization transmission where both continuous and discrete spectra are jointly modulated would therefore represent the complete system where all the degrees of freedom for modulation provided by a \ac{SMF} are exploited.

\indent In this work, we therefore demonstrate joint dual-polarization \acf{NFDM} modulation for the first time. First we discuss the framework to perform a joint \ac{INFT} operation, i.e., the operation of constructing a dual-polarization time-domain waveform with a desired dual-polarization continuous and discrete spectrum.  The steps required are described in details and the numerical algorithms that can be employed to implement the \ac{INFT} are briefly discussed and numerically characterized. By applying such \ac{INFT}, a dual-polarization joint \ac{NFDM} system is experimentally characterized.  A transmission distance of up to 3200~km is demonstrated by jointly modulating the dual-polarization signal with a 10-GBd \ac{QPSK}-modulated continuous spectrum and a 2-eigenvalue 125-MBd \ac{QPSK} discrete spectrum, for a total net line rate of 8.4~Gb/s (after \ac{FEC} overhead subtraction). The rate is mainly limited by the wide guard intervals required to fulfill the \ac{NFT} vanishing boundary conditions at the receiver, i.e., after the dispersion-induced signal broadening.

The paper is organized as follows: first in Section~\ref{sec:DPNFDM} the theoretical framework for joint \ac{NFT} and \ac{INFT} operation is provided. The discussion on the transmitter and receiver \ac{DSP} algorithms chosen for implementing the dual-polarization joint \ac{NFDM} system and the achievable digital back-to-back performance follow in Section~\ref{sec:DSP}. The experimental setup used as testbed is described in Section~\ref{sec:Setup} together with the characterization of the optical back-to-back performance. The transmission results are reported and discussed in Section~\ref{sec:Transmission} and the conclusions drawn in Section~\ref{sec:Conclusions}. 

\section{Dual-polarization joint continuous and discrete spectral modulation} 
\label{sec:DPNFDM}
In this Section, the framework for dual-polarization joint \ac{NFDM} modulation is described. First the channel model will be briefly outlined in Section~\ref{subsec:ch}, followed by the direct \ac{NFT} in Section~\ref{subsec:NFT}. Finally  Section~\ref{subsec:INFT_NCG} describes in details one of the key contributions of this work, i.e., the joint \ac{INFT} using the \ac{DT}. Throughout this section, bold symbol characters indicate vector or matrices, while their components are indicated with subscripts (without bold): e.g., $\boldsymbol\phi=(\phi_1,\dots, \phi_N)$ and  $\mathbf v=(v_1,\dots,v_N)$.

\subsection{Channel model}
\label{subsec:ch}
Let us consider a \ac{SMF} exhibiting random birefringence, and whose dispersion and nonlinear lengths are much longer than the birefringence correlation length~\cite{ManakovJETP74}. Under those conditions, the averaged \ac{MS} describes the evolution in the fiber of the two complex-envelope polarization components of a signal \mbox{$\flds[\kk] = \fldl[\kk] $, $\kk = 1,2$}~\cite{MenyukJLT06},
\begin{equation}\label{eq:MS}
  \left\{\begin{aligned}
    \pdv{\flds[1]}{\ssp}&=- \iunit\dfrac{\dispersion}{2}\pdv[2]{\flds[1]}{\ttm}+\iunit\frac{8\nonlinfact}{9}\left(|\flds[1]|^2+|\flds[2]|^2\right)\flds[1]\\
    \pdv{\flds[2]}{\ssp}&=- \iunit\dfrac{\dispersion}{2}\pdv[2]{\flds[2]}{\ttm}+\iunit\frac{8\nonlinfact}{9}\left(|\flds[1]|^2+|\flds[2]|^2\right)\flds[2]
  \end{aligned}\right.
\end{equation}
where $\ttm$ is the retarded time, $\ssp$ the distance, $\dispersion$ the \ac{GVD}, and $\nonlinfact$ the Kerr nonlinear coefficient of the fiber.

The normalized \ac{MS}~\cite{Ablowitz2004} for the anomalous dispersion  regime \mbox{($\dispersion<0$)} is
\begin{equation}\label{eq:NMS}
  \left\{\begin{aligned}
    \iunit\pdv{\nflds[1]}{\nssp}&= \pdv[2]{\nflds[1]}{t}+2\left(|\nflds[1]|^2+|\nflds[2]|^2\right)\nflds[1]\\
    \iunit\pdv{\nflds[2]}{\nssp}&= \pdv[2]{\nflds[2]}{t}+2\left(|\nflds[1]|^2+|\nflds[2]|^2\right)\nflds[2]
  \end{aligned}\right.
\end{equation}
where $\nttm$ is the normalized retarded time, and $\nssp$ the normalized distance. The equation is derived from \eqref{eq:MS} through the change of variables 
\begin{equation}\label{eq:normalization}
   \nfld[j] = \dfrac{\fld[j]}{\sqrt{P}}, \hspace{0.7cm} \nttm = \dfrac{\ttm}{T_0}, \hspace{0.7cm} \nssp = -\dfrac{\ssp}{\nftL},
\end{equation}
with $P = |\dispersion|/(\frac{8}{9}\nonlinfact T_0^2)$, $\nftL = 2 T_0^2 /|\dispersion|$, and $T_0$ is the free normalization parameter.

\subsection{Direct NFT}
\label{subsec:NFT}
The direct \ac{NFT} computes the nonlinear spectrum of the time-domain dual-polarization signal $\mathbf{q}(t) = (q_1(t), q_2(t))$. The spectrum is composed of a continuous (dispersive) part and a finite number of discrete components, which correspond to the solitonic components.
The nonlinear spectrum is defined through the  \ac{Z-S}
problem \cite{Ablowitz2004,CivelliOE18pol,ManakovJETP74} 
\begin{equation}
\matL(\nflds)\eigf=\eig \eigf
\label{eq:eigev_prob}\end{equation}
where $\matL$ is a $3\times3$ operator that depends on the  signal $\nfld$, and is given by
\begin{equation}
 \matL = 
  \begin{pmatrix} \iunit \pdv{\nttm} & -\iunit \nfld[1] & -\iunit \nfld[2] \\
          -\iunit \nflds[1]^*(\nttm) & -\iunit \pdv{\nttm} & 0 \\
          -\iunit \nflds[2]^*(\nttm) & 0 & -\iunit \pdv{\nttm} \\
  \end{pmatrix}.
\end{equation}

\begin{figure*}[!t]
\centering
\includegraphics[width=.95\linewidth]{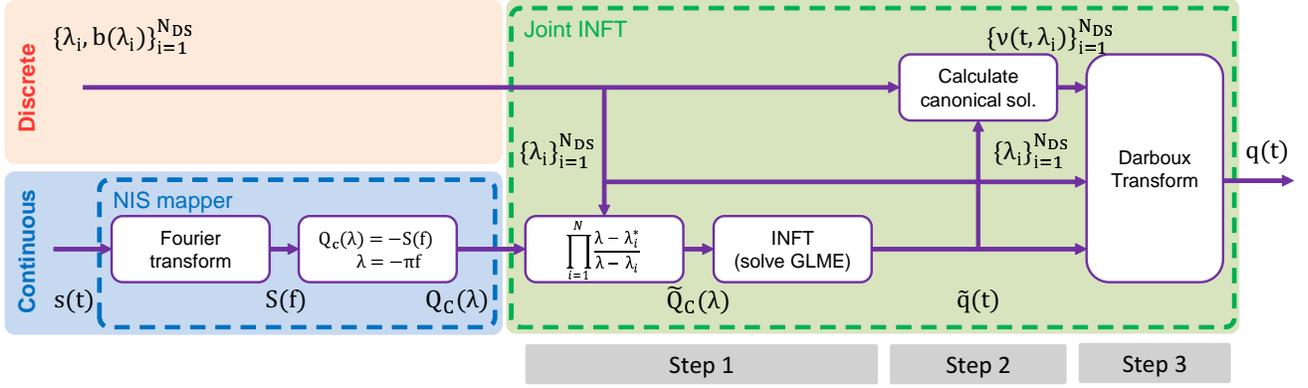}
\caption{Schematic diagram of dual-polarization joint \ac{NFDM} time-domain signal generation through a joint \ac{INFT} operation. \ul{The information is first separately encoded into discrete and continuous spectrum, through b-modulation and} \ac{NIS} \ul{ operation (via the signal Fourier transform $S(f)$), respectively. Then the three steps of the joint }\ac{INFT}\ul{ follow to obtain the time-domain waveform $q(t)$.} }
\label{fig:INFT}
\end{figure*}

The canonical solutions of \eqref{eq:eigev_prob} are the solutions $\jostns$, $\jostnconjs$, $\jostps$, and $\jostpconjs$ defined by the boundary conditions\cite{Ablowitz2004,CivelliOE18pol} 
\begin{equation}
\overbrace{
\begin{array}{cc}
{\jostnll\sim\begin{pmatrix}
1\\
0\\
0
\end{pmatrix}e^{-\iunit\eig\nttm}}, & {\jostnll[][][\bar]\sim\begin{pmatrix}
0 & 0\\
1 & 0\\
0 & 1 \\
\end{pmatrix}e^{\iunit \eig \nttm}} \end{array}
}^{\displaystyle{\text{ as }\nttm\rightarrow-\infty}}
\label{eq:asympt1}
\end{equation}
\begin{equation}
\underbrace{
\begin{array}{cc}
{\jostpll\sim\begin{pmatrix}
0 & 0 \\
1 & 0 \\
0 & 1 \\
\end{pmatrix}e^{\iunit\eig\nttm}}, & {\jostpll[][][\bar]\sim\begin{pmatrix}
1\\
0\\
0\\
\end{pmatrix}
e^{-\iunit\eig \nttm}} \end{array}
}_{\displaystyle{ \text{ as }\nttm\rightarrow+\infty.}}
\label{eq:asympt2}\end{equation}
The canonical solutions form two pairs of bases of the same subspace, consequently there exist some coefficients\textemdash known as scattering data\textemdash $\nfta\in\CC$, and
$\nftb\in\CC^{2\times1}$ such that

\begin{equation}
\jostnl=\jostpl\nftb+\jostpconjl\nfta.\label{eq:sc_data}
\end{equation}
Such scattering data can be obtained as 
\begin{subequations}
\begin{align}
\nfta &=\lim_{\nttm\rightarrow+\infty}\jostnl[1] e^{+\iunit\eig \nttm},\label{eq:scattering_data_infty_a}\\
\nftb[][1] &=\lim_{\nttm\rightarrow+\infty}\jostnl[2]e^{-\iunit\eig \nttm},\label{eq:scattering_data_infty_b}\\
\nftb[][2] &=\lim_{\nttm\rightarrow+\infty}\jostnl[3]e^{-\iunit\eig \nttm},\label{eq:scattering_data_infty_c}
\end{align}\label{eq:scattering_data_infty}
\end{subequations}
and are related to each other by
\begin{equation}
    \abs{\nfta}^2 + \abs{\nftb[][1]}^2 + \abs{\nftb[][2]}^2 = 1.\label{eq:scattering_data_relation}
\end{equation}
The continuous spectrum of the time domain signal $\nfld$ is defined as 
\begin{equation}
\cnft=\nftb/\nfta, \quad \eig\in\RR.\label{eq:rho}
\end{equation}
The discrete spectrum is made of a finite number $\nEig$ of eigenvalues $\eig[\ii]\in\CC^+$ such that $\nfta[\ii]=0$. Each eigenvalue has \ul{an associated spectral amplitude} (also referred to as norming constant) defined as
\begin{equation}
\dnft[\ii]=\nftb[\ii]/\nftaderiv[\ii],\quad\eig[i]\in\CC^+.
\end{equation}
with $\nftaderiv[i]=\frac{da(\eig)}{d\eig}|_{\eig=\eig[\ii]}$.\\

Several methods to perform the \ac{NFT} numerically are available for the single polarization scenario \cite{TuritsynOptica17,YousefiTransInfTh14,WahlsECOC17} and for the dual polarization case \cite{GaiarinOptica17,GoossensOE17,CivelliOE18pol}. In this work, the trapezoidal discretization method was used for computing both the continuous and the discrete spectrum~\cite{Arefarxiv18}.

\subsection{Inverse NFT}
\label{subsec:INFT_NCG}
The \ac{INFT} is the operation to generate a time domain signal from a given nonlinear spectrum. Several methods exist to numerically perform the \ac{INFT}  for the scalar (single polarization) \ac{NLSE} \cite{TuritsynOptica17,LeJLT15,ArefJLT18,YousefiTransInfTh14,CivelliTIWDC15}. \ul{Additionally, the theory can be extended to the} \ac{MS}~\cite{Ablowitz2004}\ul{ and a few techniques reported leading to recent numerical analysis and experimental demonstrations}: the \ac{DT} for the \ac{MS} \cite{GaiarinOptica17} for a multi-soliton signal (no continuous spectrum), the inverse Ablowitz-Ladik method for a signal with only continuous spectrum \cite{GoossensOE17}, and the \ac{NCG} method---based on the solution of the \ac{GLME}---that can be applied to a full spectrum \cite{AricoCalc11,CivelliOE18pol}. However, when the overall energy of the signal, and so of the nonlinear spectrum, is too high, the latter method may diverge~ \cite{CivelliOE18pol}.
This issue becomes relevant in optical communication when discrete eigenvalues are used for modulation \cite{CivelliOE18pol}. In this section, we describe an alternative method to generate a time domain signal from a given continuous and discrete nonlinear spectrum. Extending the concepts of \cite{ArefJLT18,Arefarxiv18} to the \ac{MS}, the method uses an \ac{INFT} algorithm (e.g., the \ac{NCG} in this work) to obtain the time-domain signal corresponding to a pre-modified continuous spectrum, and then adds the discrete eigenvalues using the \ac{DT} \cite{GaiarinOptica17,WrightAML03}. Using this approach, the aforementioned issues of \ac{NCG} can be \ul{relaxed} as the energy of the input to the \ac{NCG} is decreased by the large amount carried by the discrete spectrum, \ul{therefore shifting the energy barrier to higher energy levels}. A choice of initialization parameters is further provided to ensure that the obtained signal has the desired nonlinear spectrum, by following the approach in~\cite{Arefarxiv18,GaiarinOptica17}.  \textcolor{black}{Whereas the algorithm for single polarization in~\cite{ArefJLT18} is supported by the mathematical framework demonstrated in~\cite{Arefarxiv18}, a rigorous mathematical framework is beyond the scope of this work. The numerical accuracy of the scheme is rather confirmed a posteriori with numerical simulations (see Section~\ref{sec:DSP}) and relies on following the single-polarization framework of~\cite{ArefJLT18,Arefarxiv18} similarly to the approach of \cite{GaiarinOptica17}. Similarly to the single-polarization case, the three steps are illustrated below.}

Assume that we want to digitally compute the time domain signal $\nfld$ corresponding to the continuous spectrum $\cnft$ and the discrete spectrum $\left\{ \eig[\ii],\nftb[\ii]\right\} _{\ii=1}^{\nEig}$\textcolor{black}{, adopting $\mathbf{b}$-modulation on the discrete part as will be discussed in Section~\ref{sec:DSP}. \ul{The choice of b-modulation is not a strict requirement for the scheme presented in the following, and the approach can be easily extended to mapping techniques other than b-modulation}.}\\
The algorithm consists of three steps (illustrated in \figurename~\ref{fig:INFT}):
\begin{enumerate}

\item Use the INFT to compute the time domain signal $\nfldpredarbs(t)$
corresponding to the pre-modified continuous spectrum 
\begin{equation}
\cnftpredarb=\cnft\prod_{\ii=1}^{\nEig}\frac{\eig-\eig[\ii]}{\eig-\eig[\ii]^{*}}
\end{equation}
and with empty discrete spectrum. This can be achieved by solving the \ac{GLME} equation, for example using the \ac{NCG} method. \cite{CivelliOE18pol}. 

\item  For each eigenvalue $\eig[\ii]$ for $\ii=1,\dots,\nEig$,
obtain the solution $\darbauxsoll[][\ii]$
of the eigenvalue problem $\matL(\nfldpredarbs)\eigf=\eig[\ii]\eigf$ with boundary conditions 
\begin{equation}
\begin{cases}
\darbauxsoll[1][\ii][T]=1\\
\darbauxsoll[2][\ii][-T]=-\nftb[\ii][1]\\
\darbauxsoll[3][\ii][-T]=-\nftb[\ii][2]
\end{cases}\label{eq:boundary_cond_step2}
\end{equation}
where $\nfldpredarb=0$ for $t\notin[-T,T]$. The solution $\darbauxsoll[][\ii]$
can be obtained as 
\begin{equation}
\darbauxsoll[][\ii]=\dfrac{ \jostnl[][\ii]}{\jostns[1](T,\eig[\ii])}{}- \jostpl[][\ii]\jostps^{(2)}(-T,\eig[\ii])^{-1}\nftb[\ii],
\label{eq:v_comblin}\end{equation}
where ${ \jostnl[][\ii]}$ and ${ \jostpl[][\ii]}$ are the canonical solutions of $\matL(\nfldpredarbs)\eigf=\eig[\ii]\eigf$ (see Section \ref{subsec:NFT}) and $\jostps^{(2)}(\nttm,\eig[\ii])$ is the $2\times 2$ matrix made of the second and the third rows of $\jostpl[][\ii]$, i.e.,
\begin{equation}
{\jostps^{(2)}}(\nttm,\eig[\ii])=\begin{pmatrix}
\jostps[21] & \jostps[22]\\
\jostps[31] & \jostps[32]
\end{pmatrix}.
\end{equation}
Note that Eq.~\eqref{eq:v_comblin} is a solution of the eigenvalue problem because it is a linear combination of solutions. Furthermore,  it verifies the boundary conditions Eq.~\eqref{eq:boundary_cond_step2}.
The canonical solutions $\jostnl[][\ii]$ and $\jostpl[][\ii]$ can be found with standard methods used for the \ac{NFT} (see Section \ref{subsec:NFT}). \ul{Additionally, if modulation techniques other than b-modulation are chosen, a different mapping can be easily derived starting from Eq.}~\eqref{eq:v_comblin}.

\item Execute the \ac{DT} for the \ac{MS}, as in\cite{GaiarinOptica17,WrightAML03},
with input parameters $\nfldpredarb$, $\left\{ \eig[\ii]\right\} _{\ii=1}^{\nEig}$, and $\left\{ \darbauxsoll[][\ii]\right\} _{\ii=1}^{\nEig}$ to iteratively add the
discrete spectrum $\left\{ \eig[\ii],\nftb[\ii]\right\} _{\ii=1}^{\nEig}$
to the spectrum of $\nfldpredarb$. The $\darbauxsoll[][\ii]$ are the generic auxiliary solutions. 
The solution $\nfld$ obtained
in this manner has continuous spectrum $\cnft$
and discrete spectrum $\{\eig[\ii],\nftb[\ii]\}_{\ii=1}^{\nEig}$. 
\end{enumerate}
\textcolor{black}{This scheme can be easily verified numerically by computing the direct \ac{NFT} of the generated time-domain signal and comparing the resulting spectra with the desired ones.} 

\section{Digital signal processing}
\label{sec:DSP}

\begin{figure}[!t]
\centering
\includegraphics[width=\linewidth]{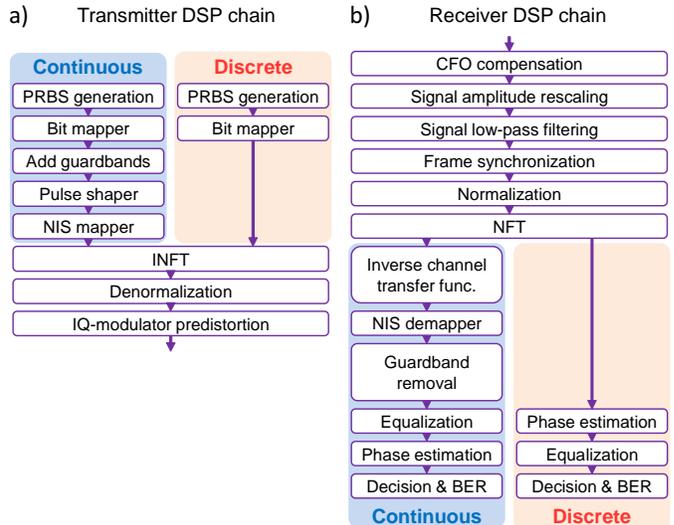}
\caption{(a) Transmitter and (b) receiver digital signal processing chains, highlighting the key operations performed on the digital waveforms.}
\label{fig:DSPchains}
\end{figure}

The \ac{DSP} chains implemented at the transmitter and receiver to properly encode the data into a digital waveform (transmitter) and extract it back (receiver) are shown in \figurename~\ref{fig:DSPchains} (a) and (b), respectively.

A \acl{PRBS} is generated at the transmitter side to be encoded in both the continuous and discrete spectra. For the continuous spectrum, after mapping the bits into 10-GBd \ac{QPSK} symbols (16 samples/symbol), guard intervals of 64~symbols are added for each 16-symbol burst, leading to an overall \ac{NFDM} symbol (burst) length of 8~ns. Such guard intervals ensure that no inter-burst interference takes place after the dispersion-induced pulse broadening at the maximum transmission distance considered in this work, i.e., 3200~km. \ul{Such long guard intervals could be decreased by adding dispersion pre-compensation at the transmitter side. This was avoided for the experimental investigation since dispersion pre-compensation would yield digital waveforms with different }\ac{PAPR} \ul{for each transmission distance, making it more challenging to compare performance at different distances}. The symbols are pulse-shaped with a raised cosine filter (roll-off of 1) and the \ac{NIS} is used to obtain the continuous spectrum as detailed in \figurename~\ref{fig:INFT}: the Fourier transform of the pulse-shaped time-domain signal is directly mapped into the continuous spectrum $\cnft$ \cite{LeOE15}. 

\begin{figure}[!t]
\centering
\includegraphics[width=\linewidth]{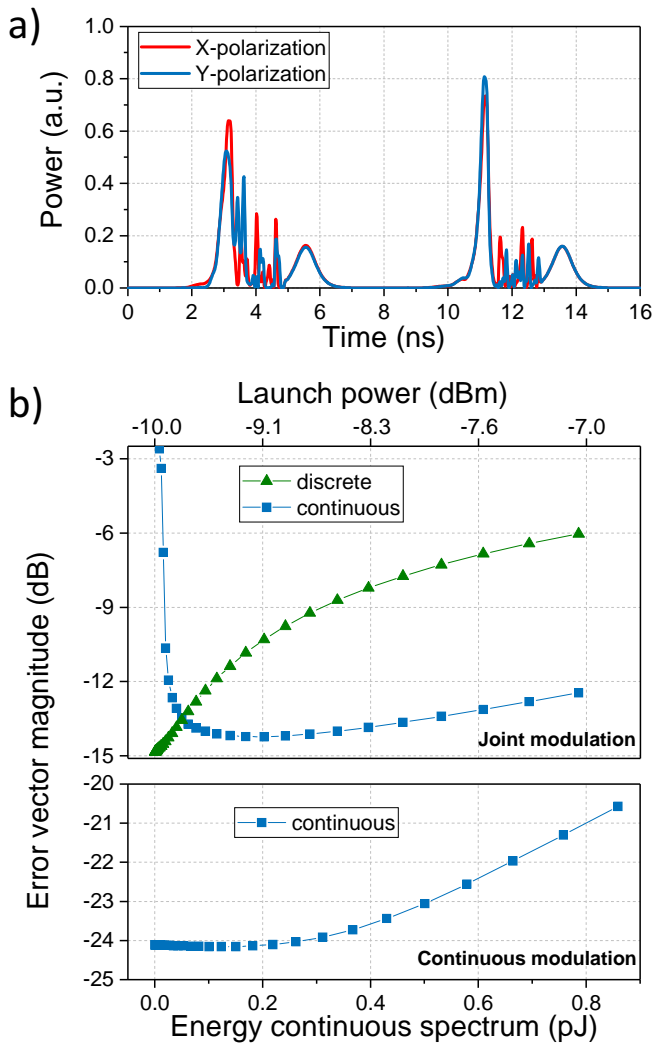}
\caption{(a) Time-domain waveforms (-9.2-dBm launch power, 2 \ac{NFDM} symbols, X and Y polarizations) showing the discrete (solitonic) components with the continuous components in between and (b) digital back-to-back performance: EVM as a function of the energy in the continuous spectrum for joint modulation (top) and continuous-only modulation (bottom).}
\label{fig:digitalPerform}
\end{figure}

The data bits to be encoded onto the discrete spectrum are also mapped to \ac{QPSK} symbols. Then, the symbols are associated to the $\nftb[i]$ ($\mathbf{b}$-modulation) corresponding to the two purely-imaginary eigenvalues, $\eig[\ii] \in \{0.3j,\, 0.6j\}$ that have been chosen. \textcolor{black}{The free time normalization parameter defined in (\ref{eq:normalization}) is chosen to be $\To = 244$~ps.} The eigenvalues themselves are not modulated, i.e., in each symbol-slot, both eigenvalues are transmitted. 
Additionally, several recent works clearly showed that modulating directly the $\nftb[i]$ ($\mathbf{b}$-modulation) of the discrete spectrum leads to a lower correlation and thus better performance than modulating $\dnft[i]$ ($\dnft[]$-modulation). In this work we therefore focus on $\mathbf{b}$-modulation for the discrete spectrum~ \cite{GuiJLT17,BuelowPTL18}.
The radii of the two \ac{QPSK} constellations have been set to $5\sqrt{2}$ and $0.05\sqrt{2}$, for the $\nftb[i]$ associated to $\eig[1]$ and $\eig[2]$, respectively. This choice leads to a temporal separation of the components of the time-domain signal associated to the \ac{NFT} coefficients $\nftb[i]$ corresponding to different eigenvalues. The separation is such that the discrete spectral components (at the transmitter output) are placed in time within the guard intervals of the continuous-spectrum burst. The one corresponding to $\nftb[1]$ can be seen in \figurename~\ref{fig:digitalPerform}(a) after the burst encoded in the continuous spectrum, whereas the waveform corresponding to $\nftb[2]$ is located on the opposite end of the symbol slot. \ul{This choice was made to avoid additional signal-dependent implementation penalties due to high} \ac{PAPR} \ul{at the transmitter side, as well as to limit the time-frequency product of the multi-soliton signal}~\cite{SpanArXiv17}. \ul{Nevertheless, continuous and discrete time-components do interact during fiber propagation.}
An \ac{INFT} operation is then performed as described in Section~\ref{subsec:INFT_NCG} to generate a time-domain waveform with the desired continuous and discrete spectra. After proper denormalization, the waveform shown in \figurename~\ref{fig:digitalPerform}(a) is obtained (signal power of -9.2~dBm with the fiber parameters of Section~\ref{sec:Setup}). The figure clearly shows the two discrete (solitonic) components with the continuous (dispersive) components in between. Finally, the waveforms are pre-distorted to account for the nonlinear transfer function of the IQ modulator by applying an $\arcsin(\cdot)$ function. Such a digital waveform can then be encoded onto an optical carrier using a standard IQ modulator, after digital-to-analog conversion, as will be described in Section~\ref{sec:Setup}. The net line rate of the generated signal is 8.4~Gb/s, taking into account the 80\%-guard intervals applied and the 7\%-\ac{HD-FEC} overhead ~\cite{AgrellIPC18,SunOE08}. 

At the receiver side, the digital waveforms are then processed by the \ac{DSP} highlighted in Fig~\ref{fig:DSPchains}(b). First, \ac{CFO} compensation is performed to remove any frequency shift due to frequency mistmatch between signal and \ac{LO}, followed by signal amplitude rescaling, low-pass filtering at twice the 20-dB signal bandwidth, and frame synchronization. The direct \ac{NFT} described in Section~\ref{subsec:NFT} is then applied to recover the continuous and discrete spectra from the time-domain waveform. The inverse transfer function of the channel $\exp(4\iunit\lambda^2\nssp)$ is first applied to the continuous part of the spectrum, followed by \ac{NIS} demapping with the opposite transformation applied at the transmitter side: first $\cnft$ is mapped into the Fourier spectrum, then a inverse Fourier transform is used to recover the time-domain waveform~\cite{LeOE15}.  The guard intervals are subsequently removed and blind radius-directed equalization is performed followed by phase estimation using digital phase-lock loop. Finally, decisions on the symbols are taken and the \ac{BER} is counted.
The steps for the demodulation of the discrete spectrum consists of first phase recovery using \ac{BPS} independently on each constellation $\nftb[i][j]$, followed by \ac{NFT}-domain equalization~\cite{GuiJLT17,GaiarinPTL18}. As the chosen eigenvalues are purely imaginary, \ac{BPS} inherently applies the ideal inverse channel transfer function, which consists of a constant phase rotation. 
After \ac{BPS}, NFT-domain equalization reduces the noise on the $\nftb[i][j]$ by exploiting the correlation between the received eigenvalues and the spectral amplitudes~\cite{GaiarinPTL18,GuiJLT17}. This equalizer enables to partially compensate for the rotation and re-scaling experienced by $\nftb[i][1]$ and $\nftb[i][2]$ due to the displacement of the eigenvalues. After equalization, decisions are taken and \ac{BER} counting is performed.

The \ac{DSP} chains and numerical algorithm have been first benchmarked in a digital back-to-back scenario where the digital waveforms before the IQ-modulator predistortion are fed directly into the receiver \ac{DSP} chain. \textcolor{black}{This analysis allows to ignore the impact of practical equipment limitations, such as \ac{ADC} and \ac{DAC} resolution, and electrical/optical noise sources, to focus on the numerical algorithms.} The resulting performance is shown in \figurename~\ref{fig:digitalPerform}(b), as a function of the energy in the continuous spectrum, comparing joint and continuous-only modulation. For the joint modulation (top figure in \figurename~\ref{fig:digitalPerform}(b)), the energy in the discrete spectrum is kept constant to fulfill the  duration-amplitude relation~\cite{YousefiTransInfTh14}. The signal quality is evaluated by calculating the \ac{EVM} as the \ac{BER} values are too low for reliable error counting~\cite{ShafikICEC06}. 

In the case of joint modulation, as the energy in the continuous spectrum increases, the performance of the continuous spectrum improves (EVM decreases) with an optimum at approx.\ 0.18~pJ (-9.2~dBm of launch power). \textcolor{black}{When the energy in the continuous spectrum approaches zero, the discrete spectrum is dominant and worsens the accuracy of the numerical algorithms for the continuous part.} 
\begin{figure*}[!t]
\centering
\includegraphics[width=.95\linewidth]{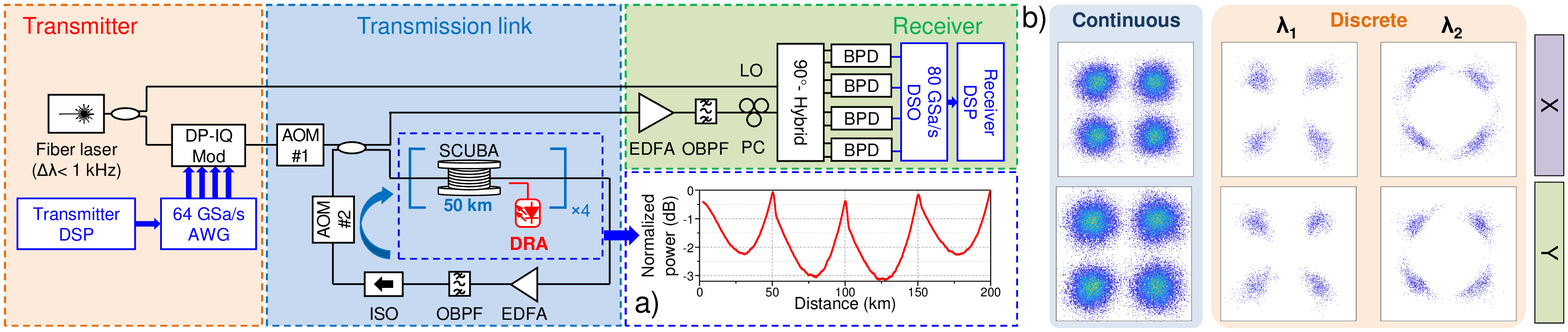}
\caption{Experimental setup for the transmission of the jointly modulated signal in a recirculating transmission loop. Insets: (a) power profile measured over a loop recirculation and (b) constellation diagrams after 2800-km transmission. }
\label{fig:Setup}
\end{figure*}
In the case of continuous-only modulation (bottom figure in \figurename~\ref{fig:digitalPerform}(b)), the performance do not degrade as the energy decreases as for the joint-modulation, thus ruling out numerical errors of the \ac{NCG} alone.\\

Beyond the optimum energy for the joint-modulation, the performance worsens rather rapidly above -9.0~dBm. A similar worsening of the performance is indeed reflected when no discrete spectrum is present. Note that, considering $\mathbf{b}$-modulation also for the continuous spectrum (instead of modulating directly $\cnft$) may provide further improvement~\cite{WahlsECOC17} \ul{even though using it in the context of joint spectral modulation may present some challenges}~\cite{GuiOE18}. \figurename~\ref{fig:digitalPerform}(b) shows also the impact of the energy in the continuous spectrum on the quality of the discrete spectrum. As the energy in the continuous spectrum is increased, \ul{the limited precision of the numerical algorithms yields a loss of orthogonality between continuous and discrete spectrum, thus decreasing the performance of the latter when the continuous components at high energy overlap with the solitons.} \ul{The impact of the time-overlap is expected to worsen the performance during transmission over a non-ideal (lossy and noisy) channel, as discussed in}~\cite{ArefECOC17}. Furthermore, the numerical precision of the \ac{INFT} is also expected to contribute to the overall worsening of the performance of the discrete spectrum.

Regardless of these limitations and the consequent energy balance between continuous and discrete spectrum, the \ac{BER} that can be estimated from the \ac{EVM} is well below $1\times 10^{-4}$, even for the highest power values considered.

\section{Experimental transmission setup}
\label{sec:Setup}

The experimental setup is shown in \figurename~\ref{fig:Setup}. The predistorted digital waveforms generated as in \figurename~\ref{fig:DSPchains}(a) are loaded into a 4-channel 64-GSamples/s \ac{AWG} driving the IQ modulator, which encodes the dual-polarization NFDM signal into an optical carrier generated by a low-linewidth ($\le$ 1~kHz) fiber laser. The same laser is used as \ac{LO} at the receiver side. 

The transmission link consists of a recirculating transmission loop based on four 50-km transmission spans with distributed Raman amplification applied to each span as in~\cite{GaiarinPTL18}. Backward pumping combined with low-loss large effective area fiber (SCUBA fiber) enables to achieve maximum power variations of approximately 3~dB across the full 200-km loop length. The power profile measured by optical time domain reflectometry is shown in inset (a) of \figurename~\ref{fig:Setup}. Loss, dispersion, and nonlinear coefficient of the transmission fiber are 0.155~dB/km, 22~ps/nm/km, and 0.6~/W/km, respectively. These values have been used for the (I)\ac{NFT} (de)normalization as discussed in Section~\ref{subsec:ch}. In addiction to the transmission fiber, the loop consists of \acp{AOM} used as optical switches, an \ac{OBPF} (0.5-nm bandwidth) which suppresses out-of-band \ac{ASE} noise, an \ac{ISO}, and an \ac{EDFA} which compensates for the power loss of all these components. 

After the chosen number of recirculation turns, the signal is received with a pre-amplified coherent receiver using four \acp{BPD} and a 80-GSamples/s \ac{DSO} acting as analog-to-digital converter. For simplicity, the signal polarization is manually aligned at the receiver input with a \ac{PC}. However, demultiplexing schemes based on training sequences have already been reported~\cite{GuiCLEO18}.
After analog-to-digital conversion, the waveforms are processed offline by the DSP discussed in Section~\ref{sec:DSP} and the performance are evaluated by bit error counting performed on more than $10^{6}$ bits, ensuring a reliable \ac{BER} above $10^{-5}$. \ul{Only }\ac{BER}\ul{ values from direct error counting are reported for experimental measurements}. In the following the transmission reach is evaluated considering the \ac{HD-FEC} threshold (\ac{BER} of $3.8\times 10^{-3}$)~\cite{SunOE08,AgrellIPC18}. Remark that the frequency-offset estimation discussed in Section~\ref{sec:DSP} is necessary due to the frequency shift introduced by the \acp{AOM} which results in self-heterodyne detection rather than homodyne.
An example of constellation diagrams for continuous and discrete spectrum is shown in inset (b) of \figurename~\ref{fig:Setup} after 2800-km transmission, illustrating the high quality of the received signals.

\begin{figure}[!t]
\centering
\includegraphics[width=\linewidth]{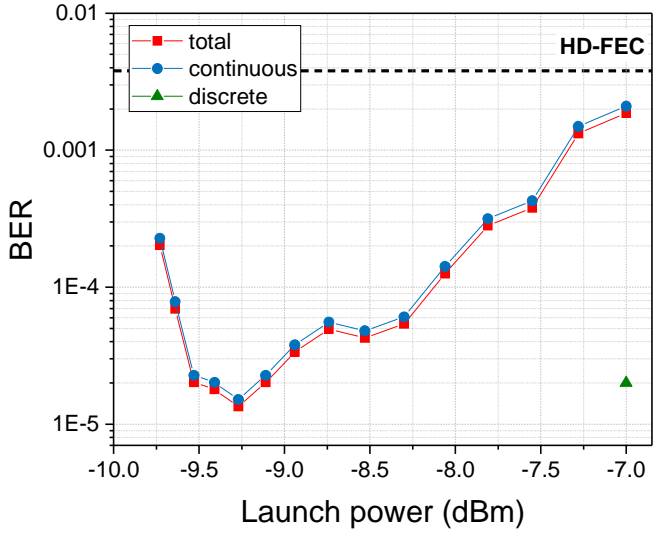}
\caption{Optical back-to-back \ac{BER} performance \ul{from direct error counting} as a function of the launch signal power.}
\label{fig:opticalBER}
\end{figure}

Before discussing the transmission results in Section~\ref{sec:Transmission}, the signal performance are evaluated in back-to-back configuration, i.e., connecting the receiver directly at the output of the IQ modulator. The results are shown in \figurename~\ref{fig:opticalBER}, distinguishing between the performance of continuous and discrete spectral components as well as showing the total \ac{BER}. The \ac{OSNR} at the output of the transmitter was approx.\ 33.8~dB for all the launch powers considered.

\ul{For these back-to-back results}, the total \ac{BER} is dominated by the \ac{BER} of the continuous spectrum \ul{as more bits are encoded in the continuous spectrum (64 bits/}\ac{NFDM} \ul{symbol) compared to the discrete spectrum (8 bits/}\ac{NFDM} \ul{symbol). Therefore, a higher error probability can be tolerated on the discrete spectrum before it starts affecting the total} \ac{BER}. As discussed for the numerical results of \figurename~\ref{fig:digitalPerform}, the \ac{BER} of the continuous spectrum improves with its increased energy, reaches an optimum and worsens due to numerical instabilities. Note  that the optimum power is the same for the digital back-to-back (see \figurename~\ref{fig:digitalPerform}(b)). The lack of variations in the optimum power hints that the dominant limitation is currently related to the numerical algorithms, whereas the impact of electrical/optical noise \ul{at transmitter and receiver}, as well as the  \ac{AWG} resolution are rather negligible.
The \ac{BER} on the discrete spectrum is also consistent with \figurename~\ref{fig:digitalPerform}(b), as errors are only detected at the highest launch power considered, -7~dBm. At such a power level, the estimated \ac{BER} in digital back-to-back was estimated to almost the same value, showing that a negligible penalty is introduced by the optical-frontends (at both transmitter and receiver) also for the discrete spectrum.

\section{Transmission performance}
\label{sec:Transmission}

\begin{figure}[!t]
\centering
\includegraphics[width=\linewidth]{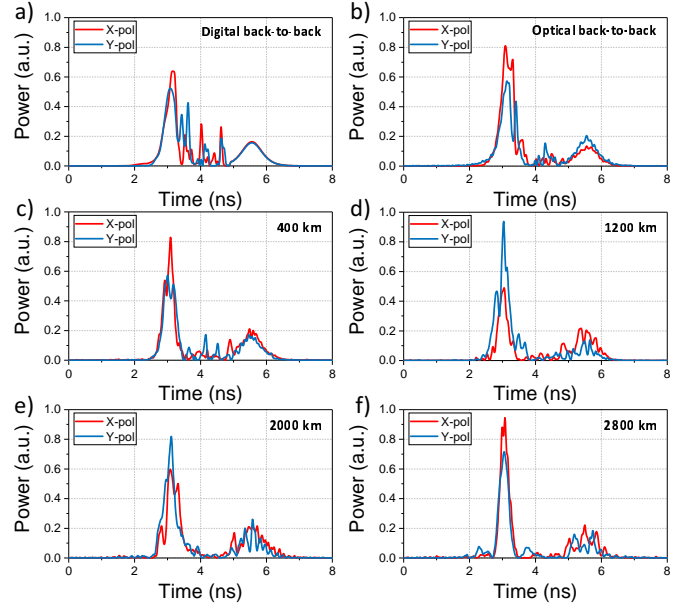}
\caption{Examples of time-domain waveforms showing one 8-ns \ac{NFDM} symbols (including guard intervals) at a fixed launch power of -9.2~dBm: (a) digital back-to-back, (b) optical back-to-back and after (c) 400-km, (d) 1200-km, (e) 2000-km, (f) and 2800-km transmission. The bit patter is not the same for the same for the different waveforms.}
\label{fig:waveforms}
\end{figure}

After having evaluated the system performance for both digital and optical back-to-back, the transmission performance is reported in this section. \figurename~\ref{fig:waveforms} shows the evolution along the fiber of one \ac{NFDM} symbol at the optimum launch power of -9.2~dBm. The waveforms clearly show the interaction in time between the discrete and continuous spectral components, whereas the guard interval size is more than sufficient to guarantee the vanishing boundary conditions required by the \ac{NFT} also at the longest transmission distances. The guard interval size could actually be reduced by pre-dispersing the waveforms at the transmitter side by half of the transmission length, i.e., by applying the inverse of the channel transfer function~\cite{TavakkolniaCLEO16}. Additionally, by tailoring the guard intervals to the desired transmission distance, the transmission rate can be maximized.
\begin{figure}[!t]
\centering
\includegraphics[width=\linewidth]{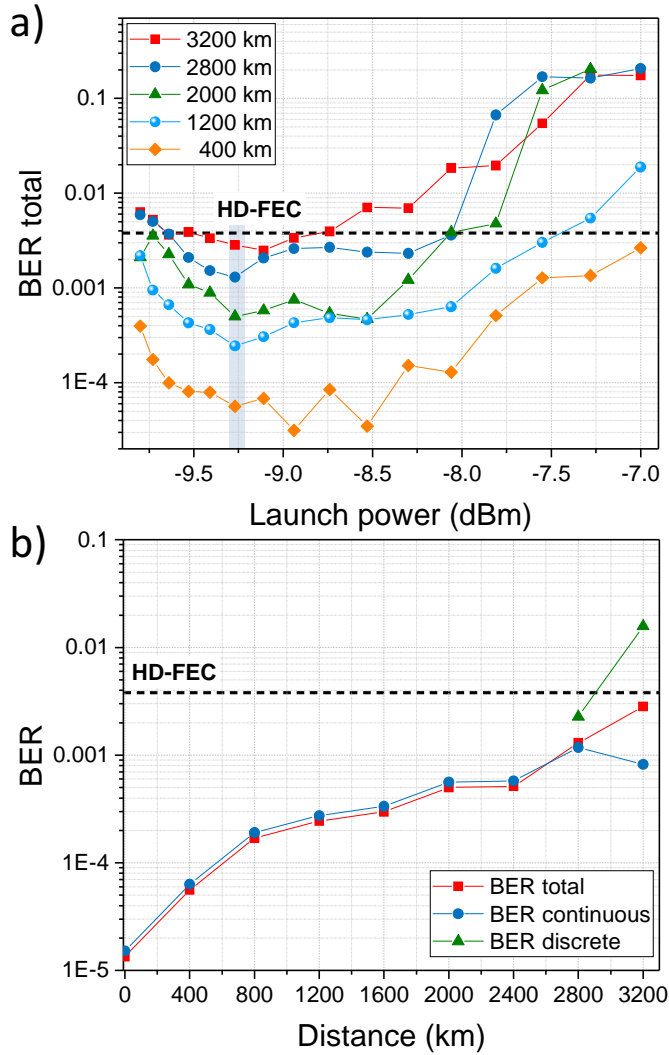}
\caption{(a) Total \ac{BER} performance as a function of the launch power for different transmission distances, and (b) total \ac{BER} and contributions from continuous and discrete spectrum as a function of the transmission distance for the optimum launch power of -9.2~dBm (shaded area in (a)).}
\label{fig:transmissionBER}
\end{figure}
The total \ac{BER} results as a function of the launch power for different transmission distances are shown in \figurename~\ref{fig:transmissionBER}(a). The curves show an optimum launch power (minimum \ac{BER}) consistent with the digital and optical back-to-back performance, i.e., $-9.2$~dBm. The \ac{BER} values after a 400-km transmission are actually in close agreement with the optical back-to-back ones in Figs.~\ref{fig:digitalPerform}(b). These results further confirm that the dominant performance limitation is currently linked to the numerical precision in performing joint \ac{INFT} and \ac{NFT} at high power values. \textcolor{black}{By increasing the numerical precision of the different steps (Section~\ref{sec:DPNFDM}) using improved numerical algorithms,} we believe the performance could be significantly improved, potentially shifting the optimum launch power to higher values~\cite{ChimmalgiArxiv18}.

The \ac{BER} as a function of the transmission distance is shown in \figurename~\ref{fig:transmissionBER}(b), for a fixed -9.2-dBm launch power. The figure highlights that \ac{BER} below the \ac{HD-FEC} threshold can be achieved for up to 3200-km transmission. Beyond 2800~km, the dominant contribution to the total \ac{BER} comes from the discrete spectral components and in particular from the largest eigenvalue $\eig[2]=0.6j$. \textcolor{black}{Improved modulation schemes} for the discrete spectral components, such as the differential modulation proposed in \cite{Spanarxiv18} or soliton detection based on matched filter\cite{ArefJLT18}, are thus expected to increase the transmission reach. Furthermore, as far as the continuous spectrum is concerned using improved detection strategies, may also provide additional performance gain~\cite{CivelliOE18detect,JonesPTL18}.  Finally, the net transmission rate may be increased by reducing the guard intervals, as mentioned above, and  pre-dispersing the signal by half of the transmission length at the transmitter side~\cite{LeNatPhot17,CivelliPTL17,TavakkolniaCLEO16}.

\section{Conclusion}
\label{sec:Conclusions}
We have introduced a framework for dual-polarization \ac{NFDM} systems which allows encoding data on both continuous and discrete spectral components. The steps to perform the joint \ac{INFT} at the transmitter side are described and numerically implemented, evaluating its performance first in a transmission-free scenario without (digital back-to-back) and with the optical front-ends (optical back-to-back) considered. The dual-polarization joint \ac{NFDM} system has then been experimentally demonstrated in a transmission scenario using distributed Raman amplification. A transmission  reach of 3200~km is achieved for a 8.4-Gb/s net rate \ac{NFDM} signal, mainly limited by the numerical implementation of the joint \ac{INFT} and \ac{NFT}, which will need to be further improved. This work demonstrates the use of all the degrees of freedom available for \ac{NFDM}-based transmission over \acp{SMF}.

\section*{Acknowledgment}
This work is supported by the European Research Council through the ERC-CoG FRECOM project (grant agreement no. 771878) and the National Council for Scientific and Technological Development (CNPq), Brazil, grant 432214/2018-6. We thank OFS fitel Denmark for providing the SCUBA fiber used in the experiment and the anonymous reviewers for their constructive feedback.

\ifCLASSOPTIONcaptionsoff
  \newpage
\fi

\end{document}